\documentclass[11pt]{article}
\usepackage{moriond,epsfig}
\usepackage{amssymb, mathrsfs}
\usepackage{epsfig, graphicx, graphics}
\usepackage{hyperref}

\def\be{\begin{equation}}
\def\ee{\end{equation}}
\def\bea{\begin{eqnarray}}
\def\eea{\end{eqnarray}}

\long\def\symbolfootnote[#1]#2{\begingroup
\def\thefootnote{\fnsymbol{footnote}}\footnote[#1]{#2}\endgroup}

\begin{document}

\pagestyle{plain}

 \begin{flushright}
{\small
 MIT-CTP-3947}
 \end{flushright} 

\vspace*{4cm}
\title{QUANTUM BOLTZMANN EQUATIONS IN RESONANT LEPTOGENESIS
\symbolfootnote[1]{Talk given at the 43$^{\rm rd}$ Rencontres de Moriond, La Thuile (Italy), 1-8 March 2008.}}

\author{A.~DE SIMONE}

\address{Center for Theoretical Physics, Massachusetts Institute of Technology, \\ Cambridge, MA 02139, USA}

\maketitle\abstracts{
The quantum Boltzmann equations relevant for leptogenesis, obtained using non-equilibrium quantum field theory, are described. They manifest memory effects leading to a time-dependent CP asymmetry which depends upon the previous history of the system. This result is particularly relevant in resonant leptogenesis where the asymmetry is generated by the decays of nearly mass-degenerate right-handed neutrinos. The impact of the non-trivial time
evolution of the CP asymmetry is discussed either in the generic resonant leptogenesis scenario or in the more 
specific Minimal Lepton Flavour Violation framework. Significant quantitative differences arise with respect to the usual approach in which the time dependence of the CP asymmetry is neglected.}

\section{Introduction}

In our universe, the difference between the number densities of baryons and anti-baryons, per entropy density, is observed to be  \cite{wmap5}
$Y_B\equiv{(n_B-n_{\bar B})/ s}=(8.84\pm 0.24) \times 10^{-11}$. This number, obtained from measurements of the Cosmic Microwave Background Radiation, is also in excellent agreement with the independent fit from Big Bang Nucleosynthesis (BBN).
Thermal leptogenesis \cite{leptogen} is a simple and well-motivated mechanism to explain this baryon asymmetry. 
The simplest implementation of this mechanism is realized by adding three right-handed (RH) Majorana neutrinos to the Standard Model (SM), {\it i.e.} the framework of type I see-saw. 
The fact that the same see-saw framework 
may simultaneously account for small neutrino masses and the baryon asymmetry of the universe makes it very attractive.
In thermal leptogenesis, the heavy RH neutrinos are produced by thermal scatterings in the early universe after inflation, and subsequently decay out of equilibrium in a lepton number and CP violating way, thus satisfying Sakharov's conditions. A lepton asymmetry then arises, which is partially converted into a baryon asymmetry by electroweak sphaleron interactions.

In the case where the RH neutrinos masses are hierarchical, successful  leptogenesis requires 
the RH neutinos to be heavier than $10^9$ GeV. Since they need to be produced after inflation, 
the reheating temperature cannot be much lower than their mass. In supersymmetric scenarios, this
may be in conflict with the upper bound on the reheating temperature necessary to avoid the overproduction of gravitinos during rehating, which may spoil the successful predictions of BBN. On the other hand, if the RH neutrinos are nearly degenerate in mass, the self-energy contribution to the CP asymmetries may be resonantly enhanced, thus making leptogenesis viable at  temperatures as low as TeV. 
This interesting situation is called ``resonant leptogenesis'' \cite{res}.

In order to precisely quantify the lepton asymmetry generated by the leptogenesis mechanism, one needs to keep track of the abundances of the particles involved in the process by solving a set of coupled Boltzmann equations. The standard calculations employ a set of semi-classical equations. However,  quantum Boltzmann equations for leptogenesis have
been recently derived \cite{dsr} (for an earlier study, see
Ref. \cite{buchmuller}), using a Green's function
technique known as Closed Time-Path  (CTP) --- or 
Schwinger-Keldysh --- formalism \cite{chou}, which provides a complete
description of  non-equilibrium phenomena in field theory.
While in the semi-classical setup every scattering
in the plasma is independent of the previous one, 
 in a full quantum approach the whole dynamical history of the
system is taken into account.
 The quantum Boltzmann
equations describe therefore a non-Markovian dynamics, manifesting
 the typical ``memory'' 
effects which are observed in quantum transport theory \cite{dan}. 
The thermalization 
rate obtained from the quantum transport theory may be 
substantially longer than the one obtained from 
the classical kinetic theory. 

Furthermore, and more importantly, the CP asymmetry turns out to be
 a function of time, even after taking the Markovian limit. 
Its value at a given
instant depends upon the previous history of the system. 
If the timescale of the  variation of the CP asymmetry is shorter than the
relaxation time of the particles abundances,
the CP asymmetry may be averaged over many scatterings
and  it reduces to its classical constant value; the solutions to the
quantum and the classical
Boltzmann equations are expected to 
differ only by terms of the order of the ratio
of the timescale of the CP asymmetry and the relaxation timescale of the
particle distributions. In thermal leptogenesis with hierarchical 
RH neutrinos this is typically the case. However, in the
resonant leptogenesis scenario, where  at least  
two RH neutrinos are almost degenerate
in mass and their   mass difference
$\Delta M$
is of the order of their decay rates, the typical timescale
to build up coherently the CP asymmetry (of the order of $1/\Delta M$)
can be larger than the timescale  
for the change of the abundance of the
RH neutrinos. Thus, in the case of resonant leptogenesis
significant differences are expected between the classical and the
quantum approach.

\section{Quantum Boltzmann equations}

The model I consider consists of the SM plus 
three RH neutrinos 
$N_i$ ($i=1,2,3$), with Majorana masses $M_1\leq M_2\leq M_3$.  
The interactions among RH neutrinos, Higgs doublets $H$, lepton doublets  
$\ell_\alpha$ and singlets $e_\alpha$  ($\alpha=e,\mu,\tau$) 
are described by the Lagrangian
\be
\mathscr{L}_{\rm int}=\lambda_{i \alpha}\,N_i\,\ell_\alpha\, H+h_\alpha\,\bar e_\alpha\,\ell_\alpha\, H^c+{1\over 2}M_i N_i^2+{\rm h.c.}\,,
\label{Lint}
\ee
with summation over repeated indices. 
In the early universe, the quantum numbers conserved by sphaleron interactions are $\Delta_\alpha=B/3-L_\alpha$, where $B, L_\alpha$ are the baryon asymmetry and the lepton asymmetry in the flavour $\alpha$, respectively.

The quantum Boltzmann equations  describing the generation of the baryon asymmetry
 are obtained using the CTP formulation of  non-equilibrium quantum field theory.
The reader is referred to Ref.~\cite{dsr} for the technical details of the calculation.  Here, I only summarize the main results.  
After taking the Markovian limit, the  equations for the number densities of RH neutrinos $Y_{N_i}$ and the asymmetries $Y_{\Delta_\alpha}$ (per entropy density) read
\bea
\frac{d Y_{N_i}}{d z} &=& - {D_i}   \left(Y_{N_i}-Y_{N_i}^{\rm eq} \right),
\label{BENi} \\ \nonumber \\
\frac{d Y_{\Delta_\alpha}}{d z} &=& - \sum_i \epsilon_{i\alpha} D_i  \left(Y_{N_i}-Y_{N_i}^{\rm eq} \right) -  
W_\alpha |A_{\alpha\alpha}| Y_{\Delta_\alpha}\,,
\label{BEDelta}
\eea
where the ratio between the mass of the lightest RH neutrino and the temperature $z=M_1/T$ plays the role of the time variable.
At equilibrium the $N_i$ number density normalized to the entropy density of the universe is given by  
$Y^{\rm eq}_{N_i}= z_i^2 \mathcal{K}_2(z_i)/(4 g_*)$, where $z_i=z \sqrt{x_i}$, $x_i= (M_i/M_1)^2$, 
 $g_*=106.75$ and $\mathcal{K}_n(z_i)$
is a modified Bessel function of the $n$-th kind.
The decay and washout terms appearing in (\ref{BENi})-(\ref{BEDelta}) are defined as
\be
D_i=K_i~ x_i~ z ~\frac{\mathcal{K}_1(z_i)}{\mathcal{K}_2(z_i)}\,\,,\qquad
W_\alpha= \sum_i \frac{1}{4} K_{i\alpha} ~\sqrt{x_i}~ \mathcal{K}_1(z_i)~ z_i^3\,,
\ee
where the washout parameters are given by the ratios between the decay rates and the Hubble parameter
\be
K_{i\alpha}={\Gamma (N_i \to \ell_\alpha \bar{H})\over H(T=M_i)}\,\,,\qquad 
K_i=\sum_\alpha K_{i\alpha}\,.
\label{Kialpha}
\ee
The form of the matrix $A$ depends on the number of lepton flavours which are effective in the dynamics of leptogenesis, and this in turn depends on the temperature at which leptogenesis takes place, which is roughly given by $M_1$. Indeed, for $M_1\gtrsim 10^{12}$ GeV all lepton flavours are not distinguishable and the one-flavour regime holds;
for $10^9 {\rm GeV} \lesssim M_1 \lesssim 10^{12} {\rm GeV}$ and $M_1 \lesssim 10^{9}$ GeV, two and three lepton flavours become effective, respectively \cite{f1,dsr4}.
For example,  in the approximation where $A$ is a diagonal matrix $A=-{\rm diag}(151/179,344/537,344/537)$,
for $M_1 \lesssim 10^{9}$GeV. The complete expressions
can be found in Refs.~\cite{f1}.
Finally, sphaleron interactions introduce a conversion factor for the final baryon asymmetry
\be
Y_B=\frac{12}{37}  \sum_\alpha Y_{\Delta_\alpha}(z\rightarrow \infty)\,.
\ee 
The key quantities controlling the production of a net lepton number 
are the CP  asymmetries in the $N_i$ decays 
\be
\epsilon_{i \alpha} \equiv \frac{
\Gamma (N_i \to \ell_\alpha \bar{H}) - \Gamma (N_i \to \bar{\ell}_\alpha H)
}{
\Gamma (N_i \to \ell_\alpha \bar{H}) + \Gamma (N_i \to \bar{\ell}_\alpha H)
}\,.
\ee

The Eqs.~(\ref{BENi})-(\ref{BEDelta}) reproduce exactly the usual Boltzmann equations obtained in the semiclassical approach, 
except for a crucial difference in the source term of (\ref{BEDelta}).
As mentioned above, the inclusion of quantum effects introduces a time dependence in the CP asymmetry
\bea  
\epsilon_{i\alpha}(z)&=& \sum_{j\neq i} \epsilon^{(j,i)}_{\alpha}\,
 m^{(i,j)}(z)\,, \\ 
m^{(i,j)}(z)&=&2 ~\sin^2\left(\frac{1}{2} \frac{M_j-M_i}{2 H(M_1)} z^2\right)
- \frac{\Gamma_j}{M_j-M_i}~\sin\left(\frac{M_j-M_i}{2 H(M_1)}z^2\right)\,,\\
\epsilon^{(j,i)}_{\alpha}&=&\frac{1}{8\pi}
\frac{\textrm{Im}\left[ \lambda_{i\alpha} \lambda^\dagger_{\alpha j} 
(\lambda \lambda^{\dagger})_{ij} \right] }
{\left(\lambda \lambda^{\dagger}\right)_{ii}}~ (g_s^{(j,i)}+g_v^{(j,i)})\,, \label{mij}\\
g_s^{(j,i)}&=&\sqrt{\frac{x_j}{x_i}} ~\frac{1}{1-\frac{x_j}{x_i}} ~\frac{1}{1+\frac{\Gamma_j^2/M_i^2}{(1-x_j/x_i)^2}}\,,\\
g_v^{(j,i)}&=&\sqrt{\frac{x_j}{x_i}} ~\left( 1- (1+\frac{x_j}{x_i})~ \ln \frac{1+x_j/x_i}{x_j/x_i} \right)\,,
\eea
where $\Gamma_j\equiv \sum_\beta \Gamma (N_j \to \ell_\beta \bar{H}) =(\lambda\lambda^\dagger)_{jj} M_{j}/(8 \pi)$ 
 is the total decay rate of the $j$-th RH neutrino,
$g_s$ and $g_v$ are the self-energy and the vertex correction functions, respectively.

In the quantum approach, the typical timescale for the variation of the CP asymmetry is 
\be
t=\frac{1}{2H(T)}=\frac{z^2}{2H(M_1)}\sim \frac{1}{M_j-M_i}=\frac{1}{\Delta M_{ji}}\,.
\ee
If the timescale for the variation of the particle abundances  $1/\Gamma_i$ is much larger than $1/\Delta M_{ji}$, the CP asymmetry will  average  to its  classical value $\bar\epsilon_{i\alpha}=
 \sum_{j\neq i} \epsilon^{(j,i)}_{\alpha}$ and no
significant quantum effect arises. On the other hand, quantum effects are expected to be sizeable if 
the timescale $1/\Delta M_{ji}$ for building up coherently the CP asymmetry  is larger than the timescale $1/\Gamma_i$  for changing the abundances. 
In other words, the oscillation frequency $\Delta M_{ji}$ has to be sufficiently smaller 
than $\Gamma_i$, so that the factors $m^{(i,j)}(z)$ do not effectively average to  one.  
Under these conditions,  the amplitude of the ``sin" term in $m^{(i,j)}(z)$ is also enhanced, 
which turns out to be a crucial effect. 

The above discussion allows one to formulate a quantitative criterion for the importance of 
quantum effects, namely  $\Delta M_{ji}  \lesssim \Gamma_i$.
This  criterion can be naturally satisfied  if RH neutrinos are nearly degenerate, such as in resonant leptogenesis  and 
in models  based on Minimal Lepton Flavour Violation.

\section{Application to resonant leptogenesis}

As anticipated in the Introduction,  resonant leptogenesis relies on the fact that the CP asymmetries
are resonantly enhanced  when the mass differences among RH neutrinos  are comparable to their decay widths. $\Delta M_{ij}\sim \Gamma_i\sim \Gamma_j$. Therefore, the  criterion for the significance of the quantum effects is satisfied and one expects the quantum Boltzmann equations to provide appreciably different results with respect to their semi-classical counterparts.

For the sake of simplicity, let us restrict here  to  the simplest  case where there are only two RH neutrinos with mass difference $\Delta M$ and similar decay rates $\Gamma_N$, in the  one-flavour approximation where $\alpha$ has a single value. In this case, the Boltzmann equation for  the lepton asymmetry in the single flavour $\alpha$  contains the CP asymmetry
\be
\epsilon_{\alpha}(z)\simeq \bar\epsilon_{\alpha}\left[2\sin^2\left({K z^2\over 4} {\Delta M\over \Gamma_N}\right)-{\Gamma_N\over \Delta M}\sin\left({K z^2\over 2} {\Delta M\over \Gamma_N}\right)\right]\,,
\label{epsz}
\ee
where $\bar\epsilon_{\alpha}= \sum_i\sum_{j\neq i} \epsilon^{(j,i)}_{\alpha}$ is the  constant value of the asymmetry in the classical limit, and $K$ is the total washout parameter.

\begin{figure}[t]
\centering
\includegraphics[scale=0.9]{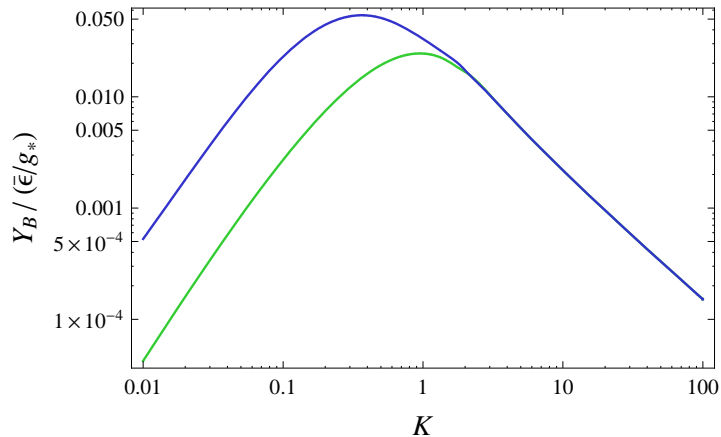}
\caption{\footnotesize The absolute value
of the final baryon asymmetry, as a function of $K$, for $\Delta M=\Gamma_{N}$. 
In blue,  the time dependence in the  CP asymmetry is included; in green, the usual 
time-independent CP asymmetry is used.}
\label{plot}
\end{figure}

The plot in Fig.~\ref{plot} shows the absolute values of the final lepton asymmetry computed with and without the time-dependent factor in (\ref{epsz}). For the strong washout regime $K\gtrsim 1$, the quantum and semi-classical methods give almost the same answers; instead, at small $K$, the discrepancy between the two approaches is sizeable, of an order of magnitude or more. This is easily understood from Eq.~(\ref{epsz}), where at large $K$ the ``sin'' functions average to constants while in the opposite case they determine an  appreciable time-dependence of the CP asymmetry.

In Ref.~\cite{dsr2}, the reader can find a more detailed study of the more general ``flavoured'' case as well as analytical approximations for the lepton asymmetries in the different possible washout regimes. They reproduce fairly well the numerical solutions of the Boltzmann equations  (\ref{BENi})-(\ref{BEDelta}).

%%%%%%%%%%%%%%%%%%%%%%%%%%%%%%%%%%%%%
%%%%%%%%%%%%%%%%%%%%%%%%%%%%%%%%%%%%

\subsection{MLFV leptogenesis}

Nearly degenerate RH neutrinos naturally arise in 
the context of models satisfying 
the hypothesis of Minimal Flavour Violation (MFV)~\cite{MFV,MLFV}.
In the quark sector, where the MFV hypothesis has been formulated first, 
the MFV ansatz states that the two quark Yukawa couplings are the only 
irreducible breaking sources of the flavour-symmetry group defined by 
the gauge-kinetic lagrangian~\cite{MFV}. In generic models satisfying this 
hypothesis, quark  Flavour Changing  Neutral Currents  are naturally suppressed 
to a level comparable to experiments and new degrees of freedom can 
naturally appear in the TeV range. The extension of the MFV hypothesis
to the lepton sector (MLFV) has been formulated in Ref.~\cite{MLFV},
where a number of possible scenarios, depending on the field content 
of the theory, have been identified. The case more similar to the quark sector and 
more interesting from the point of view of leptogenesis is the 
so-called extended field content scenario.
The lepton field content is extended by three heavy RH neutrinos
with degenerate masses at the tree level. 
The largest lepton flavour symmetry group 
of the gauge-kinetic term is $SU(3)_{\ell } \otimes SU(3)_{e} 
\otimes O(3)_{N}$ and, according to the MLFV hypothesis, it is assumed that 
this group is broken only by the charged-lepton and neutrino 
Yukawa couplings $h_\alpha$ and $\lambda_{i\alpha}$. 
In relation to leptogenesis, the key feature of this scenario is that the degeneracy of 
the RH neutrinos is lifted only by 
corrections induced by the Yukawa couplings,
so that we end up with a highly constrained version of resonant leptogenesis.

Within this setup,  the viability of leptogenesis has  been considered either in the 
one-flavour approximation~\cite{Cirigliano:2006nu} or  in the 
flavoured case~\cite{Branco:2006hz}.
However, in the light of the quantum  version of the Boltzmann equations  I discussed here,  it turned out necessary to carry out an analysis to assess the impact of the quantum effect in this
MLFV leptogenesis scenario.
It has been shown in \cite{dsr3}, both analytically and numerically, that neglecting the time dependence of the
CP asymmetry may underestimate the baryon asymmetry by several orders of magnitude when 
a  strong degeneracy among heavy RH neutrinos and small mass splittings in the light
neutrino sectors are present. This is true both when  the CP phases 
come from the RH sector and when they come entirely from the left-handed
sector  and may be identified with the low energy PMNS phases.

\section{Conclusions}

The simplest  see-saw framework, where RH neutrinos are added to the particle content of the SM, may simultaneously account for the small neutrino masses and the baryon asymmetry of the universe, through the leptogenesis mechanism. Obtaining detailed results in leptogenesis requires solving the Boltzmann equations for the abundances of the particles involved. In this talk, I have presented a set of quantum Boltzmann equations which has been derived using non-equilibrium quantum field theory. The main difference with respect to the usual semi-classical equations is that the CP asymmetry is time-dependent. A criterion to discriminate situations where one should expect quantum effects to be important has been discussed. In particular, this condition is satisfied in realistic models  such as resonant and MLFV leptogenesis. 
In these scenarios, quantum effects play a significant role
and should be taken into account.

\section*{Acknowledgments}
I am grateful to the organizers for their kind invitation. This contribution is based on work done in collaboration with A.~Riotto and with V.~Cirigliano, G.~Isidori, I.~Masina.
I acknowledge support from 
 the INFN ``Bruno Rossi'' Fellowship, from the U.S. 
Department of Energy (DoE) under contract No. DE-FG02-05ER41360, and from an EU Marie Curie Grant.

\end{document}